\documentclass[epsfig,12pt]{article}
\usepackage{epsfig}
\usepackage{graphicx}
\usepackage{hyperref}

\usepackage{array}
\usepackage{amsmath}
\usepackage{amssymb}

\newcommand{\beq}{\begin{equation}}   
\newcommand{\eeq}{\end{equation}}
\newcommand{\beqn}{\begin{eqnarray}}   
\newcommand{\eeqn}{\end{eqnarray}}

\begin{document}
\unitlength = 1mm

\def\de{\partial}
\def\Tr{ \hbox{\rm Tr}}
\def\const{\hbox {\rm const.}}  
\def\o{\over}
\def\im{\hbox{\rm Im}}
\def\re{\hbox{\rm Re}}
\def\bra{\langle}\def\ket{\rangle}
\def\Arg{\hbox {\rm Arg}}
\def\Re{\hbox {\rm Re}}
\def\Im{\hbox {\rm Im}}
\def\diag{\hbox{\rm diag}}


\def\QATOPD#1#2#3#4{{#3 \atopwithdelims#1#2 #4}}
\def\stackunder#1#2{\mathrel{\mathop{#2}\limits_{#1}}}
\def\stackreb#1#2{\mathrel{\mathop{#2}\limits_{#1}}}
\def\Tr{{\rm Tr}}
\def\res{{\rm res}}
\def\Bf#1{\mbox{\boldmath $#1$}}
\def\balpha{{\Bf\alpha}}
\def\bbeta{{\Bf\beta}}
\def\bgamma{{\Bf\gamma}}
\def\bnu{{\Bf\nu}}
\def\bmu{{\Bf\mu}}
\def\bphi{{\Bf\phi}}
\def\bPhi{{\Bf\Phi}}
\def\bomega{{\Bf\omega}}
\def\blambda{{\Bf\lambda}}
\def\brho{{\Bf\rho}}
\def\bsigma{{\bfit\sigma}}
\def\bxi{{\Bf\xi}}
\def\bbeta{{\Bf\eta}}
\def\d{\partial}
\def\der#1#2{\frac{\d{#1}}{\d{#2}}}
\def\Im{{\rm Im}}
\def\Re{{\rm Re}}
\def\rank{{\rm rank}}
\def\diag{{\rm diag}}
\def\2{{1\over 2}}
\def\ntwo{${\mathcal N}=2\;$}
\def\nfour{${\mathcal N}=4\;$}
\def\none{${\mathcal N}=1\;$}
\def\ntwot{${\mathcal N}=(2,2)\;$}
\def\ntwoo{${\mathcal N}=(0,2)\;$}
\def\x{\stackrel{\otimes}{,}}

\newcommand{\cpn}{CP$(N-1)\;$}
\newcommand{\wcpn}{wCP$_{N,\widetilde{N}}(N_f-1)\;$}
\newcommand{\wcpd}{wCP$_{\widetilde{N},N}(N_f-1)\;$}
\newcommand{\vp}{\varphi}
\newcommand{\pt}{\partial}
\newcommand{\tN}{\widetilde{N}}
\newcommand{\ve}{\varepsilon}
\renewcommand{\theequation}{\thesection.\arabic{equation}}

\newcommand{\sun}{SU$(N)\;$}

\setcounter{footnote}0

\vfill

\begin{titlepage}

\begin{flushright}
FTPI-MINN-16/19, UMN-TH-3530/16\\
\end{flushright}

\begin{center}
{  \Large \bf  
 Non-Abelian Vortex in Four Dimensions as a \\[2mm]
 Critical String on a Conifold
}

\vspace{5mm}

{\large \bf  P.~Koroteev$^{\,a,b}$,  M.~Shifman$^{\,c}$ and \bf A.~Yung$^{\,\,c,d,e}$}
\end {center}

\begin{center}

$^a${\it  Perimeter Institute for Theoretical Physics,
Waterloo, ON N2L2Y5}\\
$^b${\it  Department of Mathematics, University of California,
Davis, CA 95616}\\
$^c${\it  William I. Fine Theoretical Physics Institute,
University of Minnesota,
Minneapolis, MN 55455}\\
$^{d}${\it National Research Center ``Kurchatov Institute'', 
Petersburg Nuclear Physics Institute, Gatchina, St. Petersburg
188300, Russia}\\
$^{e}${\it  St. Petersburg State University,
 Universitetskaya nab., St. Petersburg 199034, Russia}
\end{center}

\vspace{1cm}

\begin{center}
{\large\bf Abstract}
\end{center}
Non-Abelian vortex strings   supported in  a certain four-dimensional 
\ntwo Yang-Mills theory with fundamental matter 
were shown \cite{SYcstring}
to become critical superstrings. In addition to 
translational moduli non-Abelian string under consideration carries orientational and size moduli. 
Their dynamics is described by two-dimensional sigma model 
whose target space is a tautological bundle over the complex projective space. 
For the \ntwo theory with the $U(2)$ gauge group and four fundamental hypermultiplets there are six orientational and size moduli. After combining with 
four translational moduli they form a ten-dimensional target space required for a superstring to 
be critical. For the theory in question the 
target space of the sigma model is $\mathbb{C}^2\times Y_6$, where $Y_6$ is 
a conifold.
We study closed string states which emerge in four dimensions (4D) and identify them with hadrons of the 
4D bulk \ntwo theory. It turns out that most of the states arising from the ten-dimensional graviton spectrum
are non-dynamical in 4D. We find a single dynamical massless hypermultiplet associated with the deformation 
of the complex structure of the conifold. We interpret this degree of freedom as a monopole-monopole baryon 
of the 4D theory (at strong coupling).

\vspace{2cm}

\end{titlepage}

\newpage

\tableofcontents

\newpage

\section {Introduction }
\label{intro}
\setcounter{equation}{0}

This paper builds on the previous discovery of the non-Abelian solitonic vortex string in a 
certain 4D
Yang-Mills theory shown to be critical in the strong coupling limit \cite{SYcstring}.
The results 
to be reported below are
summarized in \cite{2222}.
The particular 4D theory where non-Abelian vortex is critical is \ntwo supersymmetric QCD with
U(2) gauge group and $N_f=4$ number of quark flavors. The target space of the 2D theory on the vortex string is
 $\mathbb{C}^2\times Y_6$, where $Y_6$ is conifold.
 Analyzing the closed string spectrum we find one massless hypermultiplet associated with the deformation of the complex structure of the conifold. Then we interpret this hypermultiplet in terms of the
four-dimensional Yang-Mills theory at strong coupling.

In quantum chromodynamics Regge trajectories show almost perfect linear $J$ behavior 
($J$ stands for spin). However, in all controllable examples at weak coupling a solitonic confining string
exhibits linear behavior for the Regge trajectories only at asymptotically large spins
\cite{Yrev00,Shifman2005}.
 The reason for this is that at $J\sim 1$ the physical ``string" becomes short and thick and cannot yield
linear Regge behavior. Linear Regge trajectories at $J\sim 1$ 
have a chance to emerge only if
the string at hand
satisfies the thin-string condition \cite{SYcstring},
\beq
T\ll m^2\,,
\label{thinstring}
\eeq
where $T$ is the string tension and $m$ is a typical mass scale of the bulk fields forming the string. 
The former parameter determines the string length, while the latter determines the string width.
At weak coupling $g^2\ll 1$, where $g^2$ is the bulk coupling constant, we have  $m\sim g\sqrt{T}$. 
The thin-string condition (\ref{thinstring}) is therefore badly broken.

For most solitonic strings in four dimensions, like  the Abrikosov-Nielsen-Olesen (ANO) vortices 
\cite{ANO}, the low-energy two-dimensional effective Nambu-Goto theory on the string worldsheet
is not ultraviolet (UV) complete. To make the worldsheet theory sensible to the dimension 
of the target space one has 
to take into account higher derivative corrections \cite{PolchStrom}. Higher derivative terms
run in inverse powers of $m$ and at weak coupling blow up making the string worldsheet ``crumpled''
\cite{Polyak86}. The blow up of higher derivative terms in
the worldsheet theory corresponds to the occurrence of thick and short 
``string." 

The question weather one can find an example of a solitonic string which might produce  linear 
Regge trajectories at $J\sim 1$ was addressed and answered in \cite{SYcstring}. Such a string should satisfy the
thin-string condition (\ref{thinstring}). This condition means that higher derivative correction are 
parametrically small and can be ignored. If so the low-energy worldsheet theory should be 
UV complete. This implies the  following necessary conditions:

\begin{itemize}
\item[(i)]  The low-energy world-sheet theory must be conformally invariant;

\item[(ii)] The theory must have the critical value of the Virasoro central charge.
\end{itemize}

These conditions are satisfied by the fundamental string.

\vspace{2mm}

In \cite{SYcstring} it was shown  that (i) and (ii) above are met by non-Abelian vortex string 
\cite{HT1,ABEKY,SYmon,HT2} supported in four-dimensional 
\ntwo supersymmetric QCD with the U$(N)$ gauge group, $N_f=2N$ matter hypermultiplets  and the
Fayet-Iliopoulos (FI) parameter $\xi$. The non-Abelian part of the gauge group has vanishing $\beta$ function.

The non-Abelian vortex string is 1/2
BPS saturated and, therefore,  has \ntwot supersymmetry on its worldsheet.
In addition to translational moduli characteristic of the ANO strings,
 the non-Abelian string carries orientational  moduli, as well as size moduli if $N_f>N$.
\cite{HT1,ABEKY,SYmon,HT2}, see \cite{Trev,Jrev,SYrev,Trev2} for reviews. Their dynamics
is described by two-dimensional sigma model with 
the target space 
\beq
\mathcal{O}(-1)^{\oplus(N_f-N)}_{\mathbb{CP}^1}\,,
\label{12}
\eeq
to which we will refer to as WCP$(N,N_f-N)$ model. It has a natural description in terms of gauged linear sigma model (GLSM) \cite{W93} containing $N$ positive and $N_f-N$ negative $U(1)$ charged chiral multiplets. For $N_f=2N$
the model becomes conformal and condition $(i)$ above is satisfied. Moreover for $N=2$ the 
dimension of orientational/size moduli space is six and they can be combined with 
four translational moduli to form a ten-dimensional space required for critical superstrings\footnote{It corresponds to $\widehat{c}=\frac{c}{3}=3$}.
Thus the second condition is also satisfied \cite{SYcstring}.

Given that the necessary conditions are met, a hypothesis was put forward \cite{SYcstring} 
that this non-Abelian 
vortex string does satisfy thin-string condition (\ref{thinstring}) at strong coupling regime 
in the vicinity of a critical value of $g_c^2\sim 1$. This implies  that $m(g^2) \to \infty$ at 
$ g^2\to g_c^2$.

Moreover, a version of the string-gauge duality
for the four-dimensional bulk Yang-Mills was proposed: at weak coupling this theory is in the Higgs phase and can be 
described in terms of (s)quarks and Higgsed gauge bosons, while at strong coupling hadrons of this theory 
can be understood as string states formed on the non-Abelian vortex string.
In this paper we further explore this hypothesis by studying string theory for the critical non-Abelian vortex. 
This analysis allows us to confirm and enhance the construction \cite{SYcstring}. 

Vortices in U$(N)$ theories are topologically stable and can be realized as either closed or open 
strings. Open strings need to end on some object e.g. branes. However, there are no such objects 
in \ntwo SQCD\footnote{There is a possibility for a string to end on BPS monopoles in $\mathcal{N}=1$ 
theory which is a deformation of the \ntwo SQCD by a superpotential.}. Therefore we shall focus on the 
closed strings emerging from four dimensions and we will be able to identify closed string states 
with with hadrons of the four dimensional bulk theory. 

It is worth mentioning at this point that our solitonic vortex describes only non-perturbative states. Perturbative states, in particular massless states associated with the Higgs branch of the four-dimensional theory
(see Sec.~\ref{setup}), are present at all values of gauge couplings and are not captured by the vortex string dynamics. 

The onset of the thin-string regime (\ref{thinstring}) is determined by the
ratio $T/m^2$. While the string tension is  exactly
determined by FI parameter $\xi$,
\beq
T=2\pi \xi\,,
\label{ten}
\eeq
 there is no exact formula known for mass $m$. The latter is
a (common) mass parameter for the (s)quarks and Higgsed gauge bosons, which form long non-BPS multiplets.
Their masses receive quantum corrections (see  \cite{SYrev} and Sec. \ref{setup} below). 
Thus condition (\ref{thinstring}) can be argued for but it is problematic to rigorously prove it since we are at  strong coupling.
 We can test it, however. 
 The effective hadron four-dimensional theory which emerges
from quantization of the non-Abelian string should respect general properties of the original \ntwo theory.

We perform the following four major tests of our proposal:

\begin{enumerate}
\item[(a)] {\em  \ntwo space-time supersymmetry in 4D}. From the string side it emerges due 
to \ntwot worldsheet supersymmetry and the fact that we have only closed string states
in our theory. In fact, we will show that our non-Abelian vortex is a Type IIA superstring.

\item[(b)]   {\em Absence of 4D massless graviton}. Our original bulk theory is \ntwo QCD without gravity. 
Thus we expect that 4D massless string modes not to include graviton. 

\item[(c)] {\em Absence of unwanted massless vector multiplets.}

\item[(d)] {\em The 4D massless monopole-monopole baryon} exists only at strong coupling and cannot 
be continued
to the weak coupling, where its presence would contradict previous semiclassical analysis. 
\end{enumerate}

Note that if the Calabi-Yau manifold $Y_6$ is compact then there certainly is a massless 4D graviton 
in the spectrum\footnote{An  alternative -- massless 4D spin-2 state  with no interpretation in 
terms of 4D gravity -- is ruled 
out by the Weinberg-Witten theorem \cite{ww}.} . However, since conifold is noncompact, we do not expect  
any  massless spin-2 states appearing after the reduction to 4D, nor do they exist in the bulk 4D \ntwo theory.
We shall explicitly demonstrate that the 4D graviton 
is absent due to non-normalizability of its wave function.

Moreover, we will show that 4D massless vector multiplets associated with the Killing vectors on the conifold
 are also absent due to non-normalizability of their wave functions over the internal six-dimensional space. Massless vector multiplets have natural interpretation as gauge bosons. If they were present at
strong coupling at $g^2$ close to $g_c^2$ they would remain massless at arbitrary $g^2$, in particular, at 
weak coupling\footnote{One could avoid this conclusion if gauge fields were Higgsed at weak coupling.
However, this would require an appropriate amount of massless charged matter multiplets.}.  However, we know that
there are no massless  gauge multiplets at weak coupling in the bulk \ntwo Yang-Mills theory -- all gauge fields are Higgsed.
In particular, we will show that the 4D vector multiplet associated with deformation of 
the K\"{a}hler structure of the conifold $Y_6$ in type IIA string theory is non-dynamical. 

We will address the physical meaning of the above non-normalizability. For certain non-normalizable modes we see
that their background values should be considered as coupling constants in the 4D Yang-Mills theory\cite{GukVafaWitt}. 
For others, non-normalizability is related to instability due to the presence of the 
Higgs branch in the bulk (and associated massless states).


The paper is organized as follows. In Sec.~\ref{setup} we review physics of the \ntwo SQCD,
non-Abelian vortices and introduce a string description for these vortices. In Sec.~\ref{SUSY}
we discuss \ntwo supersymmetry on the worldsheet and show that we deal with Type IIA string.
In Sec.~\ref{graviton} we briefly review the general framework to obtain 4D states from 10D massless close string states 
like graviton and discuss the normalizability of these states. In Sec.~\ref{gij} we consider the massless
vector multiplet and hypermultiplet associated with deformations of K\"ahler and complex structures
of the conifold respectively. In Sec.~\ref{interp} we give physical interpretation of the hypermultiplet
associated with deformation of the complex structure of the conifold as a monopole-monopole baryon.
We summarize our conclusions in Sec.~\ref{conclusions}. Appendix contains
explicit expressions for the metric of resolved and deformed conifolds.

\section {Non-Abelian Vortex as a Critical Superstring }
\label{setup}
\setcounter{equation}{0}
In this section we briefly review our bulk  \ntwo  Yang-Mills, non-Abelian strings that it supports and 
the corresponding worldsheet model.

\subsection{\boldmath{\ntwo} supersymmetric Yang-Mills in 4D} 
\label{bulk}

The basic bulk model we start from  is \ntwo SQCD
 with the $U(N)$ gauge group and $N_f$ massless matter hypermultiplets. 
It is described in detail in  \cite{SYmon},  see also
 the review \cite{SYrev}.
The field content is as follows. 

The \ntwo vector multiplet
consists of the  U(1)
gauge field $A_{\mu}$ and  SU$(N)$  gauge fields $A^a_{\mu}$,
where $a=1,..., N^2-1$, as well as their Weyl fermion superpartners plus
complex scalar fields $a$, and $a^a$ and their Weyl superpartners, respectively.

The matter sector of  the U$(N)$ theory contains
 $N_f$ (s)quark hypermultiplets  each consisting
of   the complex scalar fields
$q^{kA}$ and $\widetilde{q}_{Ak}$ (squarks) and
their  fermion superpartners --- all in the fundamental representation of 
the SU$(N)$ gauge group.
Here $k=1,..., N$ is the color index
while $A$ is the flavor index, $A=1,..., N_f$. In this paper we assumed the matter mass
parameters to vanish.

In addition, we introduce the
FI parameter $\xi$ in the U(1) factor of the gauge group.
It does not break \ntwo supersymmetry.

We will consider the bulk theory with $N_f=2N$. In this case the SU$(N)$ gauge coupling does not run
since the corresponding $\beta$ function vanishes.
Note however, that the conformal invariance of the bulk theory is explicitly broken by the FI parameter.

Let us review the vacuum structure and the excitation spectrum 
of the bulk theory assuming weak coupling, $g^2\ll 1$,
where $g^2$ is the SU$(N)$ gauge coupling.
The FI term triggers the squark condensation.
 The squark vacuum expectation values (VEV's)  are  
\beqn
\langle q^{kA}\rangle &=& \sqrt{\xi}\,
\left(
\begin{array}{cccccc}
1 & \ldots & 0 & 0 & \ldots & 0\\
\ldots & \ldots & \ldots  & \ldots & \ldots & \ldots\\
0 & \ldots & 1 & 0 & \ldots & 0\\
\end{array}
\right), \qquad  \langle\bar{\widetilde{q}}^{kA}\rangle= 0,
\nonumber\\[4mm]
k&=&1,..., N\,,\qquad A=1,...,N_f\, ,
\label{qvev}
\eeqn
where we present the squark fields as matrices in the color ($k$) and flavor ($A$) indices.

The squark condensate (\ref{qvev}) results in  the spontaneous
breaking of both gauge and flavor symmetries.
A diagonal global SU$(N)$ combining the gauge SU$(N)$ and an
SU$(N)$ subgroup of the flavor SU$(N_f)$
group survives, however.  This is a well known phenomenon of color-flavor locking. 

Thus, the unbroken global symmetry of the bulk 
is 
\beq
  {\rm SU}(N)_{C+F}\times  {\rm SU}(\tN)\times {\rm U}(1)\,,
\label{c+f}
\eeq
where $$\tN=N_f-N\,.$$
Here SU$(N)_{C+F}$ represents a global unbroken color-flavor rotation, which involves the
first $N$ flavors, while the SU$(\tN)$ factor stands for the flavor rotation of the remaining
$\tN$ quarks.

Now, let us briefly discuss the perturbative excitation spectrum. 
Since
both U(1) and SU($N$) gauge groups are broken by the squark condensation, all
gauge bosons become massive. In particular, the mass of the SU$(N)$ gauge bosons
is given by
\beq
m\approx g\sqrt{\xi}
\label{mWc}
\eeq
at weak coupling.

As was already mentioned, \ntwo supersymmetry remains unbroken. In fact, with the
non-vanishing $\xi$, both the squarks and adjoint scalars  
combine  with the gauge bosons to form long \ntwo supermultiplets 
with eight real bosonic components \cite{VY}. 
All states appear in the 
representations of the unbroken global
 group (\ref{c+f}), namely, in the singlet and adjoint representations
of SU$(N)_{C+F}$,
\beq
(1,\, 1,\, 0), \quad (\textbf{Adj},\, 1,\, 0),
\label{onep}
\eeq
and in the bi-fundamental representations of $\text{SU(N)}_{C+F}\times  {\rm SU}(\tN)$
\beq
\left(\bar{\textbf{N}},\, \widetilde{\textbf{N}},\, \frac{N_f}{2\tN}\right), \quad
\left(\textbf{N},\, \widetilde{\bar{\textbf{N}}},\, -\frac{N_f}{2\tN}\right)\,.
\label{twop}
\eeq

\vspace{2mm}

The representations in (\ref{onep}) and (\ref{twop})  are labeled according to three
 factors in (\ref{c+f}). The singlet and adjoint fields are  the gauge bosons, and
 the first $N$ flavors of squarks $q^{kP}$ ($P=1,...,N$), together with their 
fermion superpartners. In particular, the mass of adjoint fields is given by Eq. (\ref{mWc}).

The physical reason behind the fact that the (s)quarks transform in the adjoint or bi-fundamental 
representations of global group
is that their color charges are screened by the condensate (\ref{qvev}) and therefore they can be considered as 
mesons.

The bi-fundamental fields (\ref{twop}) represent the (s)quarks of the type $q^{kK}$ with $K=N+1,...,N_f$.
They belong to short BPS multiplets with four real bosonic components.
These fields are massless provided that the matter mass terms vanish. In fact,  in this case the vacuum (\ref{qvev}) in which only $N$ 
first squark flavors develop VEVs is
not an isolated vacuum. Rather, it is a root of a Higgs branch on which other flavors can also develop VEVs. This Higgs branch forms a cotangent bundle to the complex Grassmannian
\begin{equation}
{\cal H} = T^*\textrm{Gr}^{\mathbb{C}}_{N_f, N}\,. 
\label{eq:HiggsbranchGr}
\end{equation}
whose real dimension is \cite{APS,MY}
\beq
{\rm dim}{\cal H}= 4N\tN.
\label{dimH}
\eeq
The above Higgs branch is non-compact and is hyper-K\"ahler \cite{SW2,APS}, therefore its metric cannot be 
modified by quantum corrections \cite{APS}. In particular, once the Higgs branch is present at weak coupling
we can continue it all the way into strong coupling. In principle, it can intersect with other 
branches if present, but it cannot disappear in the theory with vanishing matter mass parameters.
We will see below that the presence of the Higgs
branch and associated massless bi-fundamental quarks has a deep impact on non-Abelian vortex dynamics.

The Higgs branch \eqref{eq:HiggsbranchGr} has a compact base defined by the condition
\beq
\bar{\widetilde{q}}^{kA}=0\, .
\label{basehiggs}
\eeq
This is the complex Grassmannian of real dimension $2N\tN$.
The BPS vortex solutions exist only on the base of the Higgs
branch. Therefore, we will limit ourselves to the vacua which belong
to the base manifold. 

Let us comment on the  U(1) charges  in (\ref{onep}) and (\ref{twop}).
The global unbroken U(1) factor in (\ref{c+f}) acts as follows. Let us make a  
U(1)$_g$ gauge transformation on quarks $q^{kA}$ (we define the U(1) quark charge to be 1/2).
To preserve the vacuum (\ref{qvev}) we compensate it by action of the generator
\beq
\left(-\frac12,..., \, -\frac12;\, \,\,\frac{N}{2\tN},..., \frac{N}{2\tN}\right),
\label{generator}
\eeq
which belongs to flavor SU$(N_f)$. Here we separated the first $N$ and the last $\tN$ entries. 
As a result, the quarks $q^{kP}$  do not transform (hence the vacuum
(\ref{qvev}) is invariant) while the quarks $q^{kK}$ 
 acquire charges $\frac{N_f}{2\tN}$, where $P=1,..., N$ and $ K=N+1,..., N_f$. This is reflected
in (\ref{onep}) and (\ref{twop}).
 
What is usually referred to as the baryonic U(1)
symmetry is a part of the U$(N)$ gauge group in our 4D Yang-Mills. Still we can identify the unbroken U(1) factor
in (\ref{c+f}) as a ``baryonic'' U(1)$_B$ symmetry. The reason is clear: the baryonic operators
constructed  as a product of two bi-fundamental quarks
\beq
{\cal B} = \varepsilon_{KK'}\, \varepsilon_{ll'}\;q^{lK}\,q^{l'K'}\,, \quad 
{\cal \widetilde{B}} = \varepsilon^{KK'}\, \varepsilon^{ll'}\,\widetilde{q}_{Kl}\,\widetilde{q}_{K'l'}\,,\quad
K, K'=N+1,..., N_f
\label{baryonoper}
\eeq
 have the U(1)$_B$ baryonic charges 
\beq
Q_B({\cal B}) = \frac{N_f}{\tN} = 2\,, \qquad Q_B({\widetilde{\cal B}}) = -\frac{N_f}{\tN} = -2, 
\label{BB}
\eeq
where we indicated the numerical values for the case we are interested in in what follows, $ N=\tN=2$.

Certainly the physical meaning of the   baryonic charge above is not the same as, say, in 
actual QCD. As we saw above,  in our  theory bi-fundamental quarks (which can be viewed
as mesons upon Higgs screening) also carry baryonic charges. Therefore, baryons can decay into
bi-fundamental mesons. We will see example of such a behavior below.

The above analysis of the Higgs phase   assumes   weak coupling. 
What happens if we increase the coupling constant $g^2$?
In fact, the bulk theory at zero $\xi$ is invariant under S-duality which interchanges strong and weak coupling regimes \cite{ArgPlessShapiro,APS}
\beq
\tau \to \tau_D = -\frac1{\tau}, \qquad \tau = \frac{4\pi i}{g^2} + \frac{\theta_{4D}}{2\pi},
\label{tau}
\eeq
where $\theta_{4D}$ is the $\theta$-angle. Therefore, even at non-zero $\xi$ the region of $g^2\gg 1$ can be described in terms of the dual weakly coupled gauge theory.

\subsection{Non-Abelian vortex strings}
\label{vortex}

The presence of the global SU$(N)_{C+F}$ symmetry is the reason for
formation of non-Abelian flux tubes (vortex strings) \cite{HT1,ABEKY,SYmon,HT2}.
The most important feature of these vortices is the presence of orientational and size zero modes.
In \ntwo bulk theory these strings are 1/2 BPS-saturated; hence,  their
tension  is determined  exactly by the FI parameter, see (\ref{ten}).

Non-Abelian vortices confine BPS monopoles of the four dimensional theory. However, as was already mentioned, 
the monopoles cannot be attached to the string ends. In fact, in the U$(N)$ theories confined  elementary monopoles 
are junctions of two ``neighboring''  non-Abelian strings, see \cite{SYrev} and Sec.~\ref{interp} 
for a more detailed discussion.

Let us have a closer look at the effective worldsheet theory for non-Abelian vortex.
Dynamics of the translational modes (which are also present for the conventional ANO string)  
in the Polyakov formulation \cite{P81} is described by the action
\beq
S_{\rm tr} = \frac{T}{2}\,\int d^2 \sigma \sqrt{h}\, h^{kl}\d_{k}x^{\mu}\,\d_{l}x_{\mu}\,,
\label{trans}
\eeq
where $\sigma^{k}$ ($k=1,2$) are the world-sheet coordinates, $x^{\mu}$ ($\mu=1,...,4$) are 4D coordinates 
and $h=\textrm{det}(h_{kl})$ where $h_{kl}$ is the world-sheet metric which is understood as an
independent variable.

If one choose $N_f=N$ , the dynamics of the orientational zero modes on the non-Abelian vortex (they become orientational
 moduli fields 
 on the worldsheet), would be described by two-dimensional
\ntwot supersymmetric $\mathbb{CP}^{N-1}$ model which is compact \cite{HT1,ABEKY,SYmon,HT2},
see \cite{Trev,Jrev,SYrev} for reviews. Size moduli do not appear in this case.
If one adds extra quark flavors, non-Abelian vortices become semilocal.
They acquire size moduli (see the review paper \cite{AchVas} devoted to Abelian semilocal vortices).  

Non-Abelian semilocal vortices in \ntwo Yang-Mills with $N_f>N$ were studied in
\cite{HT1,HT2,SYsem,Jsem,SVY}. 
The world-sheet theory for the orientational (size)
moduli of the semilocal vortex is given by the sigma model on the tautological bundle 
over the same projective space $\mathcal{O}(-1)^{\oplus \tN}_{\mathbb{CP}^{N-1}}$ where $\tN=(N_f-N)$,
which we agreed to call $WCP(N,\tN)$. Its GLSM formulation is as follows \cite{W93}. 
One introduces two types of complex fields $n^P,\,P=1,\dots,N$ and $\rho^K,\,K=N+1,\dots, N_f$, which have $U(1)$ charges $+1$ and $-1$ respectively. The orientational moduli are described by 
 the $N$-plets $n^P$
 while the size moduli are parametrized by the $\tN$-plet
$\rho^K$. 

The effective two-dimensional theory on the worldsheet has the action
\beqn
S_{\rm or} &=& \int d^2 \sigma \sqrt{h} \left\{ h^{kl}\left(
 \widetilde{\nabla}_{k}\bar{n}_P\,\nabla_{l} \,n^{P} 
 +\nabla_{k}\bar{\rho}_K\,\widetilde{\nabla}_{l} \,\rho^K\right)
 \right.
\nonumber\\[3mm]
&+&\left.
 \frac{e^2}{2} \left(|n^{P}|^2-|\rho^K|^2 -\beta\right)^2
\right\}+\mbox{fermions}\,.
\label{wcp}
\eeqn
Since fields $n^{P}$ and $\rho^K$ have charges  $+1$ and $-1$ with respect to the gauge U(1) we have
$$ \nabla_{k}=\d_{k}-iA_{k}\,, \qquad \widetilde{\nabla}_{k}=\d_{k}+iA_{k}\,.$$
The limit $e^2\to\infty$ is implied. \footnote{A remark in passing: In fact, the world-sheet theory on the semilocal non-Abelian string is
not exactly the $WCP(N,\tN)$ model  \cite{SVY}. 
Both orientational and size moduli have logarithmically divergent norm \cite{SYsem}. After an appropriate 
infrared regularization logarithmically divergent norms  can be absorbed into the definition of 
two dimensional fields. The actual theory is called $zn$ model. Nevertheless it has the same infrared physics as the GLSM in question \cite{KSVY}.}

Coupling constant $\beta$ in (\ref{wcp}) is related to the bulk coupling via 
\beq
\beta\approx \frac{4\pi}{g^2}\, .
\label{betag}
\eeq
This formula was derived at weak coupling regime in the bulk theory \cite{ABEKY,SYmon} and 
is quasiclassical.  It is modified at strong coupling.

Note that the first (and the only) coefficient of the $\beta$ function 
$\beta_1=N-\tN$ is the same for the bulk and 
world-sheet theories. It vanishes provided $N=\tN$.

The bosonic part of the total string action for the non-Abelian vortex under consideration is 
the sum of (\ref{trans}) and (\ref{wcp}),
\beq
S= S_{\rm tr} + S_{\rm or}\,.
\label{stringaction}
\eeq

As was already mentioned, the two necessary conditions for a thin string regime are met for the non-Abelian semilocal
vortex  supported in four-dimensional \ntwo Yang-Mills theory  provided the gauge group is $U(N=2)$ and the number of quark hypermultiplets is $N_f=4$ \cite{SYcstring}. Indeed,
in the conformal gauge the translational part of the action is a free theory and therefore conformal, while
the  $\beta$ function of the the orientational (size) part  is proportional to $\beta_1=N-\tN$. Thus, the condition 
of conformality
 $\beta_1=0$ implies
\beq
N=\tN,\,\, \mbox{or} \,\, N_f=2N.
\label{confinv}
\eeq
Moreover,
the number of orientational (and size) degrees of freedom in (\ref{wcp}) is 
\beq
2(N+\tN -1)=2(2N-1),
\eeq
where we subtracted 2 because of the D-term condition (see the last line in (\ref{wcp})) and $U(1)$. Requiring that this number is equal to six gives the solution\,\footnote{See \cite{SYcstring}
for details of calculation of the Virasoro central charge for our sigma model. 
Technically there are two other pairs of $N$ and $\tN$ which formally fit our construction (vanishing beta function and vanishing Virasoro central charge): $N=1, \tN=3$ and $N=3, \tN=1$, with ratio of the $U(1)$ charges for $n^A$ and for $\rho^K$ fields being equal to $-3$ and $-1/3$ respectively. Although it is straightforward to generalize GLSM \eqref{wcp}, we cannot proceed further, since the derivation of such GLSMs as theories of dynamical vortices in \ntwo SQCD along the lines of \cite{SYsem} is not available at the moment.} $N=\tN=2$, 
$N_f=4$. 
For these values of $N$ and $\tN$
the target space of the sigma model (\ref{wcp}) is resolved conifold  \cite{W93}
\begin{equation}
Y_6=\mathcal{O}(-1)_{\mathbb{CP}^1}\oplus \mathcal{O}(-1)_{\mathbb{CP}^1}\,.
\label{eq:GLSMtarget}
\end{equation}

The global symmetry of our world-sheet sigma model (\ref{wcp}) 
\beq
 {\rm SU}(2)\times {\rm SU}(2)\times {\rm U}(1)
\label{globgroup}
\eeq
is the same as the unbroken global group of the bulk theory (\ref{c+f}) for $N=\tN=2$. 
The fields $n$ and $\rho$ 
transform in the following representations:
\beq
n:\quad (\textbf{2},\,0,\, 0), \qquad \rho:\quad (0,\,\textbf{2},\, 1)\,.
\label{repsnrho}
\eeq

\subsection{Bulk duality vs. world-sheet duality}
\label{duality}

If $\tN <N$ the bulk \ntwo Yang-Mills is asymptotically free. Its coupling constant $g^2$ is frozen at the 
scale $\sqrt{\xi}$. The theory is in the weak coupling regime if $\sqrt{\xi}\gg \Lambda$, where $\Lambda$ 
is the dynamical scale. If we make $\sqrt{\xi}\ll \Lambda$ the physics can be described by weakly coupled 
infrared-free \ntwo SQCD with the gauge group U$(\tN)\times $U$(1)^{N-\tN}$ and $N_f$ flavors of 
{\em dual} quarks \cite{SYdual}, see also \cite{SYdualrev} for a review. This bulk duality is reflected in
the world-sheet duality for the sigma model on the non-Abelian vortex. Namely, the coupling 
constant $\beta$ is reflected $\beta \to -\beta$ and the roles of $N$ orientational moduli $n^P$ and
$\tN$ size moduli $\rho^K$ are interchanged \cite{SYdual}.

In the theory at hand $\tN=N=2$ and the SU(2) gauge coupling constant does not run. However,
as was already mentioned, our bulk theory has weak-strong self-duality (\ref{tau}). This duality
should be reflected in the world-sheet model as well. Indeed, the world-sheet model (\ref{wcp}) is obviously
self-dual under the reflection of the coupling constant $\beta$,
\beq
\beta \to  \beta_D = -\beta\,.
\label{CPduality}
\eeq
Under this duality the orientational and size moduli $n^P$ and $\rho^K$ interchange.
Note, that the 4D self-dual point $g^2=4\pi$ is mapped onto the 2D self-dual point $\beta=0$.
The 2D coupling constant $\beta$ can be naturally complexified if we include the $\theta$ term in the 
action of the $\mathbb{CP}^{N-1}$ model, 
$$\beta \to \beta + i\,\frac{\theta_{2D}}{2\pi}\,.$$ 
Given the complexification of $\beta$ we expect to get a generalization of (\ref{CPduality}) to complex values of the coupling which has the same fix point $\beta=0$.


%

It was conjectured in \cite{SYcstring} that thin-string condition (\ref{thinstring}) is in fact satisfied
in this theory at strong coupling limit $g^2_c\sim 1$. 
The conjecture is equivalent
to the assumption that the mass of quarks and gauge bosons $m$ 
has a singularity  as a function of $g^2$. If we assume for simplicity that there is only one singular point, then by
symmetry, a 
natural choice is the self-dual point $\tau_c =i$ or $g^2_c =4\pi$.  This gives
\beq
m^2  \to \xi\times 
\left\{
\begin{array}{ccc}
g^2, & g^2\ll 1& \\
\infty, & g^2\to 4\pi& \\
16\pi^2/g^2,& g^2\gg 1 & \\
\end{array},
\right.
\label{msing}
\eeq
where the dependence of $m$  at small and large $g^2$ follows from the tree-level formula (\ref{mWc}) and
duality (\ref{tau}). 


Thus we expect that the  singularity of mass $m$ lies at $\beta=0$. This  is the point where the non-Abelian string 
becomes infinitely thin,  higher derivative terms can be neglected and the theory of the non-Abelian 
string reduces to (\ref{stringaction}). The point $\beta=0$ is a natural choice because at this point
  we have a regime change in the 2D sigma model  {\em per se}. 
This is the point where the resolved conifold defined by the $D$-term in
(\ref{wcp}) develops a conical singularity \cite{NVafa}.

The term ``thin string'' should be understood with care. 
As was mentioned previously,  the target space of our sigma model is  {\em non-compact}, see  \eqref{eq:GLSMtarget}. 
Since the non-compact
string moduli $\rho^K$ have the string-size interpretation one might think that at large $|\rho |$ our
 string is not thin. Note, that by the thin-string condition 
(\ref{thinstring}) we mean  that the string core  is thin, and higher-derivative corrections run 
in powers of  ${\pt^2}/{m^2}$ and are negligible.

Note that  there are massless states in the bulk theory namely bifundamental quarks (\ref{twop}) 
which give rise to the continuous spectrum.
Most of  these light modes are {\em not} localized on the string and do not participate in the string dynamics. 
The only zero modes which are localized
 (in addition to the  translational modes) are the size and the orientational modes \cite{SYsem} indicated in
(\ref{wcp}). They have  logarithmically divergent norm while other light modes are power non-normalizable in 
the infrared.   All other localized modes are massive 
with mass $\sim m$. Integrating out these massive modes leads
to higher-derivative corrections running in powers of  ${\pt^2}/{m^2}$. They are negligible if   
$m$ is large, see (\ref{thinstring}). We do {\em not} integrate out zero modes. 

\section{Type IIA Description}
\setcounter{equation}{0}
\label{SUSY}

\subsection{Vortex string and bulk supersymmetry}

In this section we discuss the space-time supersymmetry of the 
non-Abelian vortex superstring \eqref{stringaction}. 
Let us first describe the fermionic content of the worldsheet theory. 
The action 
of the translational sector
of the string in the static gauge $\sigma_1=x_0$, $\sigma_2= x_3$ can be written as a free theory
\beq
S_{\rm tr} = \frac{T}{2}\,\int d^2 x \{\d_{k}x^{i}\,\d_{k}x^{i}
 + \bar{\zeta}_L\,\pt_{R}\,\zeta_L + \bar{\zeta}_R\,\pt_{R}\,\zeta_R\},
\label{transferm}
\eeq
where the worldsheet integral in the static gauge is taken over 
$x_0$ and $x_3$, $k=0,3$, while $x^i$ are transversal translational
moduli, $i=1,2$. There are four real degrees of freedom associated with complex free fermions
$\zeta_L$ and $\zeta_R$ in the translational sector.

Note that we use the static gauge because the effective worldsheet theory for the string
was derived in the static gauge from the solitonic vortex solution of the bulk theory \cite{ABEKY,SYmon}.

The  bosonic part of the worldsheet action for orientational-size moduli (of the GLSM) 
is given by (\ref{wcp}).
The fermionic superpartners of $n^P$ and $\rho^{K}$ are  fermionic 
fields $\xi^P_{L,R}$ and $\chi^K_{L,R}$ made of left and right moving modes.
They are subject to constraint
\beq
\bar{n}_P\xi^P_{L,R} -\bar{\rho}_K\chi^K_{L,R} =0.
\eeq
These fermions are related to $n^P$ and $\rho^{K}$
via \ntwot worldsheet supersymmetry.

The total number of real degrees of freedom in the fermionic orientational-size sector is 
$4(N+\tN -1)=12$ for $N=\tN=2$.
Thus altogether we have 16 fermions in the worldsheet theory in the static gauge.
This corresponds to 32 fermions in the reparametrization invariant description 
 (which reduces to
16 fermions upon fixing a physical gauge like light-cone or static gauge). These fermions are interpreted 
as $\theta$-variables in 10D space for a closed string. 
The number of $\theta$-variables corresponds to the number 
of supercharges. This number is reduced to eight upon considering the string on a six dimensional 
Calabi-Yau manifold with SU(3) holonomy \cite{GSW}. 
Eight supercharges are required in order to have \ntwo supersymmetry in 4D space.
The rest of the 10D supersymmetry is broken by the Calabi-Yau background.

As was mentioned in the Introduction this is one of the successful tests of our picture.
4D \ntwo supersymmetry which we get on the
string side matches with \ntwo supersymmetry present in the bulk QCD from the very
beginning. Imagine that we had an open vortices  in our bulk QCD. Open strings would break
4D supersymmetry down to \none on the string side. This would contradict \ntwo supersymmetry of 
our initial theory. Fortunately we do not have open vortex strings.

\subsection{Type IIA superstring}
Given the \ntwo supersymmetry in 4D the next question to address 
is whether our vortex is described by Type IIA or
Type IIB superstring theory. To answer this question we consider 10D parity transformation. As it is well known, Type IIB string
is a chiral theory and breaks parity while Type IIA string theory is left-right symmetric and conserves 
parity \cite{GSW}.

The parity transformation acts on 4D fermions as
\beq
\psi^{\alpha} \to \bar{\widetilde{\psi}}_{\dot{\alpha}}.
\label{psiP}
\eeq
(for notations see \cite{SYsem} or \cite{SYrev}).
Explicit expressions presented in \cite{BSYhet} (in the static gauge) for profile functions 
of the fermion zero modes show
that the U(1) supertranslational and SU(2) superorientational modes are proportional to
\beqn
&&\bar{\psi}_{\dot{2}}\sim (x_1+ix_2)\zeta_L, \qquad
\bar{\psi}_{\dot{2}Pk}\sim n_P\bar{\xi}_{Lk}.
\nonumber\\
&&\bar{\widetilde{\psi}}_{\dot{1}}\sim (x_1-ix_2)\zeta_R, \qquad
\bar{\widetilde{\psi}}_{\dot{1}}^{kP}\sim -\xi^k_R\bar{n}^P.
 \eeqn
Since $x_{1,2,3}\to -x_{1,2,3}$ and $n\to-n$, $\rho \to- \rho$  under parity transformation
we have
\beq
\zeta_L\to -\bar{\zeta}_R, \quad \zeta_R\to -\bar{\zeta}_L,,\quad
\xi_R^P \to -\xi_{L}^P, \quad \chi_R^K \to -\chi_{L}^K.
\label{xiP}
\eeq
Our 2D world-sheet theory is invariant under this transformation \eqref{transferm};
 thus we conclude that
the string theory of the vortex string (\ref{stringaction}) is of Type IIA.

Certainly this result matches our expectations because we started with \ntwo supersymmetric Yang-Mills
preserving 4D parity (it is a vector-like theory).  Therefore we expect that the closed string spectrum
in this theory should respect 4D parity. 

\section {Four Dimensional Reduction}
\label{graviton}
\setcounter{equation}{0}
In this section we discuss massless states which predicts our string theory in  four dimensions.

\subsection{Generalities}

Now let us  consider Type IIA string propagating in 10D space with a non-flat metric, 
\beq
\mathbb{C}^2\times Y_6\, ,
\label{10D}
\eeq
where $Y_6$ is the non-compact target space of sigma model (\ref{wcp}) which is a resolved 
Calabi-Yau conifold \cite{NVafa}. As was argued above, we expect that 
the non-Abelian vortex becomes parametrically thin and can be described by the string action (\ref{stringaction})
at strong coupling near the self-dual point $\beta=0$. Therefore, below we assume that $\beta$ is small,
$|\beta|\ll 1$. 

Strictly speaking at small $\beta$ quantum corrections in the world-sheet sigma model blow up. In other words, we can say that at small $\beta$ the gravity approximation does not work. However, if we are interested in the massless states, we can perform the supergravity computations at large $\beta$ and the extrapolate the results to strong coupling.
The massless states in the sigma model language correspond to chiral primary operators. They are protected by
\ntwot world-sheet supersymmetry. Their masses are not lifted by quantum corrections. However,
kinetic terms (the K\"ahler potentials) can acquire corrections.

The massless 10D bosonic fields of Type IIA string theory in flat ten dimensions are graviton, dilaton and 
antisymmetric   tensor $B_{MN}$, in the NS-NS sector. In the R-R sector Type IIA string gives 
one-form and three-form \cite{Polch}.
Here $M,N=1,...,10$ are 10D indices. We start  with the massless 10D graviton and examine what states it can
produce in four dimensions. In fact, the states coming from other massless 10D fields listed above can be recovered
from \ntwo supersymmetry in 4D, see for example \cite{Louis}. We will follow the standard 
string theory method which is well developed for compact Calabi-Yau spaces \cite{GSW}. Our only novel aspect
is that that for each 4D state we have to check normalizability of its wave function over the non-compact 
$Y_6$.

Massless 10D graviton is a fluctuation of the metric  $$\delta G_{MN} = G_{MN} - G_{MN}^{(0)}$$
where $G_{MN}^{(0)}$ is the metric on (\ref{10D}) which has a block form: the flat metric for $\mathbb{R}^4$ and the
Calabi-Yau metric for the conifold (see the next sections and Appendix  for an explicit expression for this metric).

Graviton should satisfy the Lichnerowicz equation
\beq
D_A D^A \delta G_{MN} + 2R_{MANB}\delta G^{AB}=0,
\label{10DLich}
\eeq
where $D^A$ and $R_{MANB}$ are the covariant derivative and the Riemann tensor, respectively,
 calculated in the background
$ G_{MN}^{(0)}$. Here the gauge $$D_A\delta G_{N}^A -\frac12 D_N\delta G_{A}^A=0$$ is imposed. For the block form
of the metric $ G_{MN}^{(0)}$ only the six-dimensional part $R_{ijkl}$ of $R_{MANB}$ is nonvanishing while
the operator $ D_A D^A$ is given by $$ D_A D^A= \pt_{\mu}\pt^{\mu} + D_i D^i$$ where the indices $\mu,\nu =1,...,4$ 
and $i,j=1,...,6$
belong to flat 4D space and $Y_6$, respectively, and we use the 4D metric with diagonal entries $(-1,\,1,\,1,\,1)$.

\vspace{1mm}

Following a standard string theory method \cite{GSW} we look for solutions of (\ref{10DLich})
assuming the factorized form of $\delta G_{MN}$
\beq
\delta G_{\mu\nu}=\delta g_{\mu\nu}(x)\,\phi_6(y), \qquad \delta G_{\mu i}=B_{\mu}(x)\,V_i(y), 
\qquad \delta G_{ij}=\phi_4(x)\,\delta g_{ij}(y)
\label{factor}
\eeq
where $x_{\mu}$ and $y_i$ are coordinates in $\mathbb{R}^4$ and $Y_6$, respectively. Moreover,
$\delta g_{\mu\nu}(x)$, $B_{\mu}(x)$
and $\phi_4(x)$ are graviton,  vector  and  scalar fields in 4D, while $\phi_6(y)$, $V_i(y)$ and 
$\delta g_{ij}(y)$
are fields on $Y_6$. 

In order for the fields $\delta g_{\mu\nu}(x)$, $B_{\mu}(x)$
and $\phi_4(x)$ to be dynamical in 4D the fields $\phi_6(y)$, $V_i(y)$ and $\delta g_{ij}(y)$ 
should have finite norm
when integrated over the six-dimensional internal space $Y_6$. Otherwise, the 4D fields come with infinite
kinetic energy and are not dynamical \cite{GukVafaWitt}. They just decouple, and this is very important.

Symbolically the Lichnerowicz equation (\ref{10DLich}) can be written as
\beq
(\pt_{\mu}\pt^{\mu} + \Delta_6)\,g_4(x)g_6(y)=0,
\label{symbol}
\eeq
where $\Delta_6$ is the two-derivative operator from (\ref{10DLich}) reduced to $Y_6$, while
$g_4(x)g_6(y)$ symbolically denotes the factorization form (\ref{factor}). If we expand
$g_6$ in eighenfunctions, 
\beq
-\Delta_6 g_6(y)=\lambda_6 g_6(y)\,,
\label{lambda}
\eeq
the eighenvalues $\lambda_6$ will play the role of the mass squared of the 4D states. 

Since our 
conifold is asymptotically flat $g_6$ for $\lambda_6>0$ behaves as a plane wave at large $y_i^2$ and 
is non-normalizable. Thus we are looking for massless 4D states with $\lambda_6 =0$
\beq
 -\Delta_6 g_6(y)=0.
\label{eighen}
\eeq
Solutions of this equation for Calabi-Yau manifolds
are given by elements of Dolbeault cohomology $H^{(p,q)}(Y_6)$, where 
$(p,q)$ denotes numbers of holomorphic and anti-holomorphic indices in the form. The dimensions of these spaces 
$h^{(p,q)}$ are called Hodge numbers for a given $Y_6$.

\subsection{4D graviton}

For 4D graviton $g_{\mu\nu}(x)$ in (\ref{factor}) equation (\ref{eighen}) takes the form
\beq
- D_i D^i \phi_6 =  -D_i \pt^i\phi_6 =0\,.
\label{4Dgraveq}
\eeq
It has only one solution 
\beq
\phi_6(y)={\rm const}\,.
\label{gravsol}
\eeq
For a compact Calabi-Yau space this is expressed as $h^{(0,0)}=1$ and leads to the presence of a single
graviton in 4D. For the conifold under consideration the solution (\ref{gravsol}) has infinite norm on $Y_6$, so there is 
no 4D graviton in our theory.

This result is expected and most welcome. As was already mentioned, the original \ntwo Yang-Mills theory in 
four dimensions had no gravity and, therefore, we do not expect 4D graviton to appear as a closed 
string state. The 
result above  is a non-trivial check of our approach and, in particular, of the validity of the
main conjecture of the thin-string regime for vortex string (\ref{thinstring}).

The non-normalizability of wave function (\ref{gravsol}), besides graviton, rules out also 
other  4D states of the \ntwo 
gravitational and tensor multiplets: vector field, dilaton,  antisymmetric tensor and two scalars
coming from 10D three-form.

Note also, that even if we ``forgot'' about the GSO projection the tachyon would be absent 
in 4D anyway  due to non-normalizability of (\ref{gravsol}).

\subsection{Killing vectors}

Consider now the second option in (\ref{factor}): 10D graviton $\delta G_{\mu i}$ gives rise 
to a vector field in 4D. This possibility is related to the presence of continuous symmetries
on $Y_6$. Our conifold $Y_6$ has  a global symmetry, so we expect to have seven Killing vectors
associated with the generators of (\ref{globgroup}).

The Killing vectors obey the following equation:
\beq
D_iV^m_j + D_jV^m_i =0\,, \qquad m=1,...,7\,.
\label{Killingeq}
\eeq
For the Calabi-Yau manifold it then follows that $V_i$ should satisfy Eq. (\ref{eighen}) which
reads
\beq
D_jD^j V_i^m=0\, .
\eeq

Being integrated by parts over compact Calabi-Yau spaces this equation implies that $V_i$ is a 
covariantly constant vector $D_jV_i=0$. Such vectors are incompatible with the SU(3) holonomy.
This leads to the conclusion that there are no global continuous symmetries on compact Calabi-Yau manifolds
\cite{GSW}.

For non-compact $Y_6$ this conclusion can be avoided and we expect presence of seven Killing vectors associated 
with the symmetry (\ref{globgroup}). However, it is easy to see that $V_i^m$ produced by rotations of coordinates 
$y_i$ by the generators of  (\ref{globgroup}) do not fall-off at large $y_i^2$ (where the $Y_6$ metric tends to flat). Thus, they 
are non-normalizable, and the associated  4D vector fields $B_{\mu}(x)$ are absent. 

This result also matches our expectations. Vector fields $B_{\mu}(x)$ naturally have interpretation of gauge
 fields. Their presence would mean that we have low energy gauge group (\ref{globgroup}) in 4D.
However as we explained in Sec.~\ref{bulk} symmetry (\ref{globgroup}) is a global unbroken group of 
our bulk \ntwo QCD. It is not gauged. Therefore the presence of gauge fields $B_{\mu}(x)$ would
lead to inconsistency of our picture. Happily they are absent.

Moreover, as was noted in Sec. \ref{intro}, massless gauge fields, if present at strong coupling,
could be continued all the way to the weak coupling domain. Then their presence would contradict the 
quasiclassical
analysis of Sec.~\ref{bulk}, where it is shown that we do not have massless gauge 
multiplets at weak coupling.

\subsection{Physical nature of non-normalizable modes}
\label{non-norm}

If we were studying the fundamental string on a non-compact Calabi-Yau space, we would conclude that
string propagates in the full 10D space and 4D subspace of it has no special role. However, our string
is a solitonic vortex in 4D gauge theory. Clearly we have to interpret string states as states living in 
this 4D  theory. Most of string states are not localized near the 4D subspace and from 4D perspective
represent non-normalizable states. What is the physical nature of these non-normalizable modes, in particular
those  we found above?

 One option is that 
non-normalizable modes, being non-dynamical, correspond to the
coupling constants of 4D theory \cite{GukVafaWitt}. One example of this is the 4D graviton considered above.
It comes with the infinite kinetic term; hence, the 4D metric cannot fluctuate. 
It is fixed to be flat and
can be viewed as a fixed background  
rather than a dynamical field. In other words, the 4D ``Planck mass'' is infinite in our theory.

Another example is the 4D gauge fields $B_{\mu}(x)$ associated with the Killing vectors. 
As was noted above, they
correspond to gauging of the global bulk symmetry (\ref{c+f}) which, if present, would contradict
consistency of our picture. However, these gauge fields also come with the infinite kinetic terms, which means that
the gauge coupling constants of these fields are in fact zero. This confirms that the symmetry (\ref{c+f}) is
global rather than local.

The most straightforward example of  this situation will be discussed in Sec. \ref{gij}. 
We will see that the coupling constant $\beta$
is  a non-normalizable modulus of the 4D theory. 

There are also non-normalizable massive 4D states associated with continuous spectrum
of (\ref{lambda}).
We interpret these modes as follows.
For these modes the associated
integrals over $Y_6$ are divergent at large $y^i$'s. Large $y^i$ mean large $n^P$ and $\rho^K$, see
(\ref{wcp}). In particular, $\rho^K$ have size moduli interpretation; they represent long-range tails of the
non-Abelian vortex in the directions orthogonal to the string axis. The very presence of these long-range 
tails (and logarithmic divergence of orientational and size zero modes \cite{SYsem}) are related 
to the presence of the Higgs branch (\ref{dimH}) and associated massless bi-fundamental quarks (\ref{twop}).

We see that the wave functions of non-normalizable states are saturated at large distances from the 
vortex string axis  in four dimensions.
  Therefore, these states are {\em not} localized on the string. The infinite norm of these 
states is interpreted as an instability.
These states are massive and therefore unstable. Namely, they decay into massless bi-fundamental quarks. 

As we already mentioned in the Introduction the vortex string 
of \cite{SYcstring}  is conceptually different in comparison 
 with fundamental string 
theory. In the theory of fundamental string  
{\em all} states present in four dimensions  are string states.
The string theory for vortex strings of \cite{SYcstring} is slightly different. The string 
states should describe 
only non-perturbative physics at strong coupling, such as mesons and baryons. The perturbative 
states seen at week coupling are not described by this theory.
 In particular, the Higgs branch (and associated massless bi-fundamental quarks)
 found at weak coupling can be continued to the strong coupling. It can
intersect other branches, but cannot disappear (for quarks with the vanishing  mass terms)
\cite{APS}.

\section {Deformations of the Conifold Metric}
\label{gij}
\setcounter{equation}{0}

In this section we consider the last option in (\ref{factor}), namely 4D scalar fields associated with 
deformations of the conifold metric $\delta g_{ij}(y)$. Eq. (\ref{eighen}) in this case reduces
to the Lichnerowicz equation on $Y_6$, namely
\beq
D_k D^k \delta g_{ij} + 2R_{ikjl}\delta g^{kl}=0.
\label{6DLich}
\eeq
Solutions of this equation for the Calabi-Yau spaces reduce to  deformations of the K\"ahler form or deformations
of complex structure \cite{NVafa,GukVafaWitt}. For a generic Calabi-Yau manifold the numbers of these deformations are given by 
$h^{(1,1)}$ and $h^{(1,2)}$, respectively. Before describing these deformations we will briefly
review  conifold geometry.

\subsection{Conifold}

The target space of the sigma model (\ref{wcp}) is defined by the $D$-term condition
\beq
|n^P|^2-|\rho^K|^2 = \beta\,
\label{Fterm}
\eeq 
and  the U(1)  phase is gauged away.
We can construct the U(1) gauge invariant  variables to be referred to as ``mesonic,"
\beq
w^{PK}= n^P \rho^K.
\label{w}
\eeq
In terms of these variables the condition (\ref{Fterm}) can be written as
\beq
{\rm det}\, w^{PK} =0\, ,
\eeq
or, alternatively, 
\beq
\sum_{\alpha =1}^{4} w_{\alpha}^2 =0,
\label{coni}
\eeq
where $$w^{PK}=\sigma_{\alpha}^{PK}w_{\alpha}$$ and $\sigma$ matrices are  chosen $(1,-i\tau^a)$, $a=1,2,3$.
Equation (\ref{coni}) defines the conifold, which is a cone whose section is $S_2\times S_3$.

At $\beta =0$ this conifold develops a conical singularity and both $S_2$ and $S_3$ shrink to zero.
It has the K\"ahler Ricci-flat metric and represents a non-compact Calabi-Yau manifold \cite{Candel,W93,NVafa}. 
The explicit form of this metric is \cite{Candel}
\beq 
ds^2=dr^2 + \frac{r^2}{6}(ds_1^2+ds_2^2) +\frac{r^2}{9}ds_3^2 ,
\label{conmet}
\eeq
where 
\beqn
&& ds_1^2= d\theta_1^2 + (\sin{\theta_1})^2 d\varphi_1^2\,,\\[1mm]
\nonumber\\
&& ds_2^2= d\theta_2^2 + (\sin{\theta_2})^2 d\varphi_2^2\,,\\[1mm]
\nonumber\\
&& ds_3^2= (d\psi  + \cos{\theta_1}d\varphi_1+ \cos{\theta_2}d\varphi_2)^2\,.
\label{angles}
\eeqn
Here $r$ is the radial coordinate on the cone while the angles above are defined at $0\le \theta_{1,2}<\pi$,
$0\le \varphi_{1,2}<2\pi$, $0\le \psi<4\pi$.

\vspace{2mm}

The volume integral associated with this metric is 
\beq
({\rm Vol})_{Y_6} = \frac{1}{108}\int r^5 dr d\psi  d\theta_1 d\varphi_1 d\theta_2 d\varphi_2 
\sin{\theta_1} \sin{\theta_2}\,.
\label{Vol}
\eeq
We can introduce another radial coordinate, $$\widetilde{r}^2 =\sum_{\alpha =1}^{4} |w_{\alpha}|^2\,.$$ It is related to
$r$ in (\ref{conmet}) via \cite{Candel}
\beq
r^2 = \frac32 \,\widetilde{r}^{4/3}\,.
\label{rtilder}
\eeq

The conifold singularity can be smoothen in two different ways: by deformation of the K\"ahler form or deformation of the 
complex structure. The first option is called resolved conifold and amounts to introducing the  non-zero
$\beta$ in (\ref{Fterm}). This resolution preserves K\"ahler structure and Ricci-flatness of the metric. 
If we put $\rho^K=0$ in (\ref{Fterm}) we get $\mathbb{CP}^1$ model with target space
 $S^2$ of radius $\sqrt{\beta}$. The explicit metric for resolved conifold can be found in 
\cite{Candel,Zayas,Klebanov}, see also Appendix B.

If $\beta=0$ there is another option -- deformation of the complex structure. It also preserves the K\"ahler 
property and Ricci-flatness of the metric of the conifold. This is called ``deformed conifold."
It  is defined by deformation of Eq.~(\ref{coni}), namely,   
\beq
\sum_{\alpha =1}^{4} w_{\alpha}^2 = b\,,
\label{deformedconi}
\eeq
where $b$ is a complex number.
Now if we take the radial coordinate $\widetilde{r}=0$ the $S_3$ does not shrinks to zero, its size is 
determined by
$b$. The explicit metric on the deformed conifold is presented in \cite{Candel,Ohta,KlebStrass},
see   Appendix B.

\subsection{The K\"ahler structure deformations}

Consider the 4D scalar field $\beta(x)$ associated with deformation of the K\"ahler form of the 
conifold $\beta$, see (\ref{Fterm}). The effective action for this field is
\beq
S(\beta) = T\int d^4x \,h_{\beta}(\pt_{\mu}\beta)^2,
\label{Sbeta}
\eeq
where the metric $h_{\beta}(\beta)$ is given by the normalization integral over the conifold $Y_6$,
\beq
h_{\beta} = \int d^6 y \sqrt{g} g^{li}\left(\frac{\pt}{\pt \beta} g_{ij}\right)
g^{jk}\left(\frac{\pt}{\pt \beta} g_{kl}\right)\,.
\label{hbetagen}
\eeq
Here $g_{ij}(\beta)$ is the resolved conifold metric, while $g$ is its determinant. Using the explicit 
 expression for the resolved conifold metric (\ref{resconmet}) we 	find 
\beq
g^{li}\left(\frac{\pt}{\pt \beta} g_{ij}\right)
g^{jk}\left(\frac{\pt}{\pt \beta} g_{kl}\right)= \frac{90}{r^4}
\eeq
to the leading approximation at small $\beta$.
Taking into account the volume integral (\ref{Vol}) we arrive at the following $\beta$ normalization integral:
\beq
h_{\beta}= (4\pi)^3\,\frac56\int d r\,r =\infty\,.
\label{hbeta}
\eeq
It is seen that the $\beta$ normalization integral is quadratically divergent in the infrared.
Thus, the scalar 4D $\beta(x)$ decouples in the bulk QCD, it is not represented by a localized state.

As was already mentioned, $\beta$ can be naturally complexified, see Sec.~\ref{duality}. On the string 
theory side the imaginary part of $\beta$ comes from 10D antisymmetric tensor. Moreover,
in Type IIA superstring the complex scalar  $\beta$ is a part of \ntwo massless vector
multiplet which also includes 4D vector field coming from the 10D three-form (see \cite{Louis} for a review).
All fields of this 4D massless vector multiplet are non-dynamical because of their infinite norm on $Y_6$.

Much in the same way as in the case of massless vector multiplets associated with the Killing vectors the 
absence of the vector $\beta$ multiplet matches our expectations. Indeed,
 massless gauge fields, if present at strong coupling,
could be continued all the way up to the weak coupling domain where their presence would contradict the 
quasiclassical analysis of Sec.~\ref{bulk}.

As was explained in Sec.~\ref{non-norm}, non-normalizable modes can be interpreted as (frozen) 
coupling constants in the 4D bulk theory. 
The $\beta$ field is the most straightforward example of this, since the 2D coupling $\beta$ is
known to be related to the 4D coupling.

\subsection{Complex structure deformations}
\label{csd}

Now let us focus on the singular point $\beta=0$. At this self-dual value of the coupling constant 
there is different deformation
of the conifold metric satisfying (\ref{6DLich}). Namely, the deformation of the complex structure 
(\ref{deformedconi}) induced by the complex modulus $b$. The effective action for this field is
\beq
S(b) = T\int d^4x \,h_{b}|\pt_{\mu} b|^2,
\label{Sb}
\eeq
where the metric $h_{b}(b)$ is given by the normalization integral over the conifold $Y_6$,
\beq
h_{b} = \int d^6 y \sqrt{g} g^{li}\left(\frac{\pt}{\pt b} g_{ij}\right)
g^{jk}\left(\frac{\pt}{\pt \bar{b}} g_{kl}\right),
\label{hbgen}
\eeq
Here $g_{ij}(b)$ is the deformed conifold metric.

We will calculate $h_b$ below using two distinct methods. The first one follows the general framework
developed in \cite{GukVafaWitt}.\footnote{We are very grateful to Cumrun Vafa for illuminating
communications and for bringing our attention to this paper.}

Using the constraint (\ref{deformedconi}) we can nominate, say, $w_2$, $w_3$ and $w_4$ as independent variables.
Then the volume form of the $Y_6$ conifold can be written as 
\beq
({\rm Vol})_{Y_6} \sim \int \left| \frac{dw_2 dw_3 dw_4}{w_1}\right|^2\,.
\label{VolV}
\eeq
The metric (\ref{hbgen}) can be expressed as 
\beq
h_b \sim \frac{\pt}{\pt b}\frac{\pt}{\pt \bar{b}}\,\int \left| \frac{dw_2 dw_3 dw_4}{w_1}\right|^2\,, 
\eeq
(see \cite{Candelas2}).
Calculating the derivatives under the constraint  (\ref{deformedconi}) we arrive at
\beq
h_b \sim \int \frac{d \widetilde{r}}{\widetilde{r}} \sim \log{\frac{\widetilde{r}^2_{\rm max}}{|b|}}\,,
\label{logV}
\eeq
where the logarithmic integral at small distances is cut off by the minimal size of  $S_3$ which is 
equal to $|b|$. 

Now let us verify this result by explicit calculations.  Starting from the explicit 
 expression for the deformed conifold metric (\ref{defconmet}) we obtain (to the leading order in $b$)
\beq
g^{li}\left(\frac{\pt}{\pt b} g_{ij}\right)
g^{jk}\left(\frac{\pt}{\pt \bar{b}} g_{kl}\right)= \frac{(\sin{\psi})^2}{\widetilde{r}^4},
\eeq
where $\widetilde{r}$ is given by (\ref{rtilder}).
Substituting this into the volume integral (\ref{Vol}) and using the relation (\ref{rtilder}) we finally get
\beq
h_b = (4\pi)^3\,\frac{4}{3}\, \log{\frac{\widetilde{r}^2_{\rm max}}{|b|}}\,.
\label{log}
\eeq

It is seen that the norm of the field $b(x)$ is logarithmically divergent in the infrared.
The modes with logarithmically divergent norm are on the borderline between normalizable 
and non-normalizable modes. Usually
such states considered as ``localized'' on the string. We follow this rule. In our framework
(vortex string vs string theory) we can
 relate this logarithmic behavior with the marginal stability of the $b$ state, see Sec. \ref{interp}.
In fact, this mode is localized on the string in the same sense as the orientational and size zero modes 
are localized on the vortex solution in the  bulk theory:
they also have logarithmically divergent norm in the infrared in 4D space \cite{SYsem}.

The upper bound in (\ref{logV}) can be related to the (infinite) size $L$  of $\mathbb{R}^4$. Noting\,\footnote{See Sec.~\ref{non-norm}
for a more detailed explanation.} that
$\widetilde{r}_{\rm max} \sim |n_{\rm max}\rho_{\rm max}| \sim \xi L^2$  we finally get
\beq
h_b = (4\pi)^3\,\frac{4}{3}\,\log{\frac{\xi^2 L^4}{|b|}} .
\label{logf}
\eeq

In Type IIA superstring the complex scalar associated with deformations of the complex structure of the Calabi-Yau
space enters, in fact, 
as a 4D hypermultiplet. Thus our 4D scalar $b$ is a part of a hypermultiplet. Another complex scalar
$\widetilde{b}$ comes from the 10D three-form (see \cite{Louis} for a review). Together they form the bosonic
component of the 4D \ntwo hypermultiplet. Thus we expect that the bosonic part of the full effective action for the
 $b$ hypermultiplet takes the SU(2)$_{R}$ invariant form,
\beq
S(b) = T\int d^4x \left\{|\pt_{\mu} b|^2 +|\pt_{\mu} \widetilde{b}|^2 \right\}\,
\log{\frac{T^4 L^8}{|b|^2+| \widetilde{b}|^2}}\,,
\label{Sbtb}
\eeq
where we absorb the constant in front of the logarithm term in (\ref{logf}) into  field normalization.
The
fields $b$ and $\widetilde{b}$ being massless can develop VEVs. Thus we have a new Higgs branch with the
metric determined by the logarithmic factor in (\ref{Sbtb}). This branch develops only at the self-dual value of the
coupling constant $g^2=4\pi$. 

The logarithmic metric in (\ref{Sbtb}) in principle can receive both perturbative and 
non-perturbative quantum corrections. However, for \ntwo  theory the non-renormalization
theorem of \cite{APS} forbids the  dependence of the Higgs branch metric  on the 4D coupling 
constant $g^2$.
Since the 2D coupling $\beta$ is related to $g^2$ we expect that the logarithmic metric in (\ref{Sbtb})
will stay intact.

To conclude this section we would like to stress that the presence of the 
new ``non-perturbative'' Higgs branch
 at a single point  $g^2=4\pi$ at strong coupling is another successful evidence for the validity of our picture. 
Indeed, a hypermultiplet is a BPS state. Were it present in some interval of $\tau$ 
at strong coupling it could be continued all the way up  to weak coupling  where its presence would 
contradict\,\footnote{In principle, one can avoid this 
conclusion if other massless BPS states are present. Together they can combine into massive non-BPS
multiplet.}
 the quasiclassical analysis, see Sec.~\ref{bulk}. 

\section {Physical Interpretation of String States}
\label{interp}
\setcounter{equation}{0}

In this section we reveal a physical interpretation of the $b$ state as a monopole-monopole baryon.

\subsection{String states at weak coupling}
\label{ssawc}

\begin{figure}
\epsfxsize=10cm
\centerline{\epsfbox{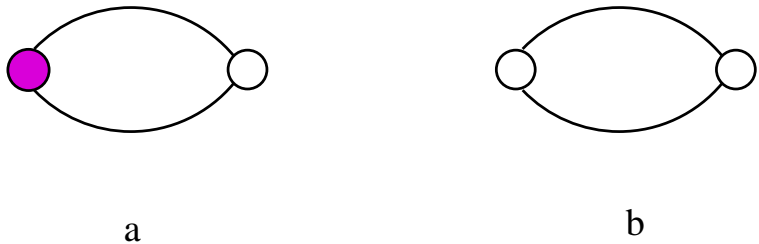}}
\caption{\small  a. Monopole-antimonopole  stringy meson. b. Monopole-monopole stringy baryon.
Open and closed circles denote the monopole and antimonopole, respectively. }
\label{mmesbar}
\end{figure}

Consider first the weak coupling region $g^2\ll 1$ in  \ntwo SQCD. Since squarks develop condensates (\ref{qvev}) non-Abelian vortices confine monopoles. As was already mentioned,  confined elementary 
monopoles
are in fact junctions of two distinct elementary  non-Abelian strings \cite{T,SYmon,HT2}. As a result,
in the bulk SQCD  we have 
monopole-antimonopole mesons in which the monopole and antimonopole are connected by two confining strings,
see Fig~\ref{mmesbar}a. In the U$(N)$  gauge theory we can have baryons  appearing as  a closed 
necklace configurations \cite{SYrev}. For the U(2) gauge group this necklace configuration
consists of two monopoles, see Fig.~\ref{mmesbar}b.

Moreover, monopoles acquire quantum numbers with respect to the global symmetry group (\ref{c+f}). To see this
note that in the worldsheet
theory on the vortex string confined monopole is seen as a kink interpolating between two different vacua
(which are distinct elementary non-Abelian strings) of the corresponding 2D sigma model \cite{T,SYmon,HT2}. On the other hand we know that the sigma model kinks at strong coupling are described by $n^P$ and $\rho^K$ fields \cite{W79,HoVa}
(for the sigma model described by \eqref{wcp} it was shown in \cite{SYtorkink}) and therefore transform in 
the fundamental representations\,\footnote{Strictly speaking 
to make  both bulk monopoles and world-sheet kinks well defined as  localized objects 
we should introduce a infrared regularization,
say, a small quark mass term. When we take the limit of the zero quark masses, the kinks become massless
and smeared all over the closed string. However their global quantum numbers stay intact.} of 
non-Abelian factors in (\ref{c+f}).

As a result monopole-antimonopole mesons and baryons in our case can be singlets or triplets
of both SU(2) global groups in (\ref{c+f}), as well as in the bi-fundamental representations. With respect
to baryonic U(1)$_B$ symmetry in (\ref{c+f}) the mesons at hand have charges $Q_{B}({\rm meson})=0,1$ while 
baryons can have charges
\beq
Q_{B}({\rm baryon})=0,\,1,\,2\,,
\label{Bbaryons}
\eeq
 see (\ref{repsnrho}). All these non-perturbative stringy states are heavy, with mass of the
order of $\sqrt{\xi}$, and therefore can decay into screened quarks which are lighter and, eventually, into
massless bi-fundamental screened quarks (\ref{twop}).

\subsection{Monopole-monopole baryon}

Now we pass to the self-dual point $\beta=0$ in strong coupling region. We will 
show that the $b$ state of the string
associated with the deformation of the complex structure of the deformed conifold 
can be interpreted as a baryon
constructed from two monopoles, see Fig. 1b. From Eq.~(\ref{deformedconi}) we see that the complex 
parameter $b$ (which is promoted to a 4D scalar field) is singlet with respect to two SU(2) factors
of the global world-sheet group (\ref{globgroup}). What about its baryonic charge? Since
\beq
w_{\alpha}= \frac12\, {\rm Tr}\left[(\bar{\sigma}_{\alpha})_{ KP}\,n^P\rho^K\right]
\label{eq:kinkbaryon}
\eeq
we see that the $b$ state transforms as 
\beq
(1,\,1,\,2),
\label{brep}
\eeq
where we used  (\ref{twop}) and (\ref{deformedconi}). In particular it has baryon charge $Q_B(b)=2$.

Since the worldsheet and the bulk global symmetries
are isomorphic we are lead to the conclusion that the massless $b$ hypermultiplet 
is a monopole-monopole baryon
with the quantum numbers (\ref{brep}) under symmetry \eqref{globgroup}.

We have observed that at infinite coupling of the two dimensional theory ($\beta=0$) a new `exotic' Higgs branch opens up, which is parameterized by the VEV of the hypermultiplet of the effective string compactification. This branch emanates only from that locus and does not exist at nonzero $\beta$.
Being massless this state is marginally stable at $\beta=0$ and can decay into pair of 
massless bi-fundamental
quarks in the singlet channel with the same baryon charge $Q_{B}=2$, see (\ref{baryonoper}). The 
$b$ hypermultiplet does not exist at non-zero $\beta$. 
One way to 
interpret this fact
in terms of bulk SQCD is as follows.
The $b$ hypermultiplet may have a ``wall of 
marginal stability" in the complex $\beta$ plane --  a closed loop 
shrunk to a single point $\beta=0$. Outside this point the $b$ hypermultiplet does not exist
as a stable state, 
 while at this point it is  marginally stable. 

This interpretation is supported by logarithmic divergence of the norm of the $b$ state 
kinetic term (\ref{Sbtb}), which in turn suggests that the $b$ state is only marginally stable. 
Detailed studies of how this can happen and how the $b$ hypermultiplet interacts with massless 
bi-fundamental quarks is left for future work.


\section {Conclusions }
\label{conclusions}
\setcounter{equation}{0}

In this paper we studied the massless spectrum produced by closed non-Abelian vortex string
in \ntwo QCD with U(2) gauge group and $N_f=4$ flavors of quark multiplets. We interpreted 4D 
closed string 
states as a hadrons of the bulk QCD. Most of the string states turns out to be non-dynamical due to
non-compactness of the six dimensional internal Calabi-Yau space $Y_6$. In particular, we showed the 
absence
of 4D graviton and unwanted vector fields in full accord with expected properties of \ntwo bulk QCD.
We found one massless 4D hypermultiplet associated with deformations of the complex structure 
of the conifold $Y_6$.  This state is present only at the self-dual point $g^2=4\pi$. We interpreted 
it as a baryon constructed from two monopoles connected 
by confining strings, see Fig.~\ref{mmesbar}b. 

We expect that this massless hypermultiplet is the lowest state of the whole  Regge trajectory
of states with higher spins in 4D. Since 4D space is flat we expect this Regge trajectory to be linear
with respect to spin $J$. The explicit construction of this Regge trajectory is left for a future work.

Let us make some comments to connect our results with other developments in string theory. 
Non-Abelian vortices appear as D2 branes extended along the finite interval between NS5 branes 
and D3 branes. The length of this interval is proportional to the FI parameter, which gives 
the string tension \cite{HT1,HT2}. 
In some other examples within the AdS/CFT framework the solitionic vortices turn out to be D-branes or D-strings wrapping some compact cycles \cite{Kleb1,Klebanov,Kleb2}. Yet, to the best of our knowledge, in the current  literature  solitonic strings so far  have not been treated as fundamental superstrings.

In the present paper (and in \cite{SYcstring}) neither did we assume the presence of the ten-dimensional space-time and fundamental strings or D-branes, nor used any holographic duality. Instead our starting point is a four-dimensional \ntwo supersymmetric QCD. Certainly this theory can be realized as a low-energy limit 
of the fundamental string theory with D branes or via geometric engineering. However, we do not assume this construction from the the beginning since, our starting basic  bulk theory {\em per se} is well defined.

Then we explored the case  $N_f=2N$ in \ntwo SQCD and found that it supports 1/2 BPS non-Abelian vortex strings. If $N=2$ the worldsheet theory on this vortex has ten real moduli which can be interpreted as coordinates on the target space $\mathbb{R}^4\times Y_6$ of the two-dimensional sigma model. This supersymmetric sigma model describes critical superstring.

Our theory predicts non-perturbative hadronic states of the original SQCD at strong coupling (at $\beta =0$).
The tension of the vortex sting is fixed by the 4D Fayet-Iliopoulos  term $\xi$, which is a scale for ``strong interactions'', not the {\em bona fide} Planck scale.
In a sense, we returned to the early days of  string theory and tried to obtain (supersymmetric) hadrons as closed string excitations of a solitonic SQCD string. It turns out that in a  proper setup it is possible. 
 
Within our approach we certainly should not think of the solitonic vortex string \cite{SYcstring} as 
of a D brane since we do not have any supergravity and D branes to begin 
with.\footnote{We are deeply indebted 
to Igor Klebanov for rising this 
issue, bringing our attention to Refs. \cite{Kleb1,Klebanov,Kleb2}, and suggesting that 
there might be a string theory S duality which relates D string and fundamental string.} However, 
it would be stimulating
 to find a possible connections between the  results reported in \cite{SYcstring,2222}  
and  the literature on solitonic strings engineered in string theory. Presumably one can see 
the spectrum of light states which we described in this work by applying some string dualities.

\section*{Acknowledgments}

The authors are grateful to Nathan Berkovits,   Alexander Gorsky, Igor Klebanov, Zohar Komargodski,
Andrei Mikhailov  and Cumrun Vafa  for very useful and 
stimulating discussions and communications.
This work  is supported in part by DOE grant DE-SC0011842. 
The work of A.Y. was  supported by William I. Fine Theoretical Physics Institute  of the  University 
of Minnesota, and by Russian State Grant for
Scientific Schools RSGSS-657512010.2. The work of A.Y. was supported by Russian Scientific Foundation 
under Grant No. 14-22-00281. P.K. would also like to thank W. Fine Institute for Theoretical Physics at University of Minnesota for kind hospitality during his visit, where part of his work was done. The research of P.K. was supported in part by the Perimeter Institute for Theoretical Physics. Research at Perimeter Institute is supported by the Government of Canada through Industry Canada and by the Province of Ontario through the Ministry of Economic Development and Innovation.

\addcontentsline{toc}{section}{Appendix}

%

\section*{Appendix. \\Metrics of  resolved and deformed conifolds}

\renewcommand{\theequation}{A.\arabic{equation}}
\setcounter{equation}{0}

The K\"ahler, Ricci flat metric on the resolved conifold has the form \cite{Candel,Zayas,Klebanov}
\beq 
ds^2=\kappa(r)^{-1}\,dr^2 + \frac{r^2}{6}ds_1^2+ \frac16(r^2 + 6\beta)\,ds_2^2 
+\kappa(r)\,\frac{r^2}{9}ds_3^2 \,,
\label{resconmet}
\eeq
where the angle differentials are defined in (\ref{angles}), while function $\kappa(r)$ is equal to
\beq
\kappa(r) = \frac{r^2 +9\beta}{r^2+ 6\beta}\,.
\label{kappa}
\eeq

Consider now the metric on the deformed conifold.
The  deformation  (\ref{deformedconi}) preserves K\"ahler structure and Ricci flatness of the   
conifold metric.
The metric of the deformed conifold  has the form \cite{Candel,Ohta,KlebStrass}
\beq 
ds^2= |b|^{2/3}\,K(u)\left\{ \frac{(\sinh{u})^3}{3(\sinh{2 u}-2 u)}\left(d u^2 + ds_3^2\right)
+ \frac{\cosh{u}}{4}(ds_1^2+ds_2^2) +\frac{1}{2}ds_4^2 \right\},\\[2mm]
\label{defconmet}
\eeq
where angle differentials are defined in (\ref{angles}), while
\beq
ds_4^2=\sin{\psi}(\sin{\theta_1}d\theta_2 d\varphi_1 +\sin{\theta_2}d\theta_1 d\varphi_2)
+\cos{\psi}(d\theta_1 d\theta_2-\sin{\theta_1}\sin{\theta_2}d\varphi_1d\varphi_2).
\eeq
Here 
\beq
K(u)=\frac{(\sinh{2 u}-2 u)^{1/3}}{2^{1/3}\sinh{u}}
\eeq and the radial variable $u$ is defined as
\beq
\widetilde{r}^2=|b|\,\cosh{u}.
\eeq

\end{document}